# Nonvolatile optical switching of surface metallicity in 1*T*-$TaSe_2$

Junde Liu[1,2,†], Liwen Su[3,4,†], Pei Liu[3,4,†], Hui Liu[1,†], Mojun Pan[1], Yuchong Zhang[1], Famin Chen[1], Yueqian Chen[3,4], Zhaoyang Xie[3,4], Stefan Mathias[2], Tianping Ying[1,*], Lin Hu[3,4,*], Tian Qian[1,*], Xun Shi[3,4,*], Yugui Yao[3,4,*]

[1]*Beijing National Laboratory for Condensed Matter Physics and Institute of Physics, Chinese Academy of Sciences, Beijing, China*

[2]*I. Physikalisches Institut, Georg-August-Universität Göttingen, Göttingen, Germany*

[3]*Centre for Quantum Physics, Key Laboratory of Advanced Optoelectronic Quantum Architecture and Measurement (MOE), School of Physics, Beijing Institute of Technology, Beijing, China*

[4]*Beijing Key Lab of Nanophotonics and Ultrafine Optoelectronic Systems, Beijing Institute of Technology, Beijing, China*

[†]These authors contributed equally to this work.

[*]Corresponding author. Email: ying@iphy.ac.cn; hulin@bit.edu.cn; tqian@iphy.ac.cn; shixun@bit.edu.cn; ygyao@bit.edu.cn

## Abstract

The surface of a quantum material provides a distinct electronic environment in which reconstruction can stabilize states absent from the bulk. However, whether light can persistently reconfigure this surface-bulk electronic contrast remains unclear. Here we demonstrate nonvolatile and reversible optical insulator-to-metal switching at the surface of the charge density wave (CDW) material 1*T*-$TaSe_2$, whose bulk is metallic. Angle-resolved photoemission spectroscopy reveals that a single femtosecond pulse creates a persistent metallic surface state. Heating or optical pulse trains restore the insulating state. Comparison of the measured band structures with first-principles calculations associates the two electronic states with distinct near-surface CDW stacking configurations that modify interlayer-mediated electron hopping. Layer-resolved calculations further show insulating outer layers above a metallic interior in the equilibrium configuration, whereas metallic states reach the outermost layer in the configuration associated with the photoinduced state. Our results establish persistent optical control of surface metallicity and open a route to reconfiguring the effective surface-bulk electronic boundary in layered quantum materials.

## Main Text

### Introduction

Surfaces of quantum materials can host low-energy electronic states distinct from those in the bulk[1–3], underpinning phenomena such as topological states[4], unconventional superconductivity[5,6] and correlated electronic states[7]. Such surface-bulk differentiation can define an effective electronic boundary beneath the physical surface that governs how electronic wavefunctions extend and hybridize across the material[8,9]. These states have typically been engineered through stacking configuration, twist-angle design and heterostructure assembly[3,10–12], but the resulting electronic architecture is generally fixed during fabrication. Whether the surface-bulk electronic contrast can be persistently and reversibly reconfigured within a single material remains largely unexplored.

Ultrafast optical excitation provides a nonequilibrium route to controlling electronic states, enabling transient manipulation of superconducting, charge-ordered, magnetic and topological phases[13–16]. A strong excitation can also drive systems across kinetic barriers into long-lived metastable or hidden states, extending optical control beyond transient responses[17–20]. Yet persistent optical control has predominantly targeted bulk order parameters and phase transformations[21–25]. At surfaces, the electronic structure depends on local reconstruction, screening and coupling to the underlying bulk, and photo-induced configurations commonly relax as the excitation energy dissipates[26,27]. Achieving nonvolatile and reversible switching of surface electronic states would therefore open a route to reconfiguring the effective surface-bulk electronic boundary.

The layered CDW material 1*T*-$TaSe_2$ provides a natural setting to explore this possibility. It crystallizes in a quasi-two-dimensional structure of 1*T*-type, which is further reconstructed to form a commensurate star-of-David (SOD) CDW below ~ 470 K[28]. Bulk transport experiments indicate that metallic behavior persists down to low temperatures, whereas surface-sensitive techniques such as angle-resolved photoemission spectroscopy (ARPES) and scanning tunnelling microscopy (STM) reveal an energy gap below ~ 260 K[29,30], indicating an insulating surface above a metallic interior. Previous studies further show that the surface electronic structure is highly sensitive to interlayer configurations[31], where possible stacking faults can lead to the emergence of distinct metallic and insulating states[32–34]. These findings establish 1*T*-$TaSe_2$ as a suitable platform for exploring reversible optical control of distinct surface electronic states.

In this work, we demonstrate nonvolatile and reversible optical insulator-to-metal switching at the surface of 1*T*-$TaSe_2$, as schematically summarized in Fig. 1. A single femtosecond control pulse induces a persistent change in the ARPES spectra, characterized by the emergence of a Ta 5*d* band crossing the Fermi level and a pronounced bandwidth increase in its in-plane dispersion. The photo-induced metallic state can be erased thermally or optically, enabling bidirectional control of the surface electronic state. Comparison with first-principles calculations associates the equilibrium and photo-induced states with distinct A-surface and L-surface configurations as defined in Figs. 1e and 1g, respectively, which are the two lowest-energy surface configurations considered (Fig. 1d). Layer-resolved calculations show that the A configuration has gapped outer layers above a metallic interior, whereas the L configuration supports metallic states in both the interior and the outermost layer. Our results suggest that changes in near-surface stacking reconfigure the surface electronic state and the effective electronic boundary between the surface and bulk.

**Results**

We first tracked the evolution of the electronic band structure accompanying the nonvolatile optical switching between the surface-insulating and metallic states in 1*T*-$TaSe_2$. At 68 K, the equilibrium ARPES spectrum measured along the Γ-M direction using a He-Iα light source exhibits a well-defined band gap near the Fermi level and a CDW gap at k ~ -0.7/Å (Fig. 2a), which is consistent with previous ARPES measurements[28,29,35]. Upon excitation by a single intense femtosecond laser pulse (1.2 eV, 80 fs, fluence ~ 1 $mJ/cm^2$), the system undergoes a pronounced and persistent reconstruction of its low-energy band structure as shown in Fig. 2b. A systematic investigation by varying the laser fluence and pulse number reveals a threshold fluence of approximately 0.7 $mJ/cm^2$, above which the nonvolatile switching occurs (see Fig. S2 in Supplementary Information). The most striking change following this transition is the collapse of the surface-insulating gap and the emergence of the Ta 5*d* band that crosses the Fermi level, establishing a metallic phase. Importantly, the low-energy Ta 5*d* band exhibits a substantially enhanced in-plane dispersion, as highlighted by the second-derivative plots (Figs. 2c and 2d) and further visualized by the corresponding EDCs (Figs. 2e and 2f). To quantify this evolution, we compared the extracted low-energy dispersions of the insulating and metallic states shown in Fig. 2g. The significantly broadened dispersion in the photo-induced metallic state indicates enhanced electronic itinerancy and a substantial modification of the effective low-energy hopping.

To explore the microscopic origin of this electronic reconstruction, we performed DFT+U calculations across different surface-stacking configurations, motivated by the strong sensitivity of the electronic structure to CDW stacking in this material[31,36]. The A- and L-surface configurations are found to be energetically favourable among other structures (see Fig. S4 in Supplementary Information), and their calculated dispersions reproduce the key experimental features, including the dispersions and overall distributions (Figs. 2h and 2i). Specifically, the A-surface calculation captures both the surface-insulating gap and the CDW gaps observed in the equilibrium state, whereas the L-surface configuration reproduces the metallic band crossing the Fermi level, its enlarged in-plane dispersion and the reduced CDW gap in the photo-induced state. To highlight the associated band renormalization, we constructed ARPES difference maps by subtracting the insulating-state spectrum from the metallic-state spectrum (Figs. 2j and 2k). Intensity gains (red) and losses (blue) reveal the redistribution of spectral weight between the two states. The calculated dispersions for the A- and L-surface configurations closely track the blue and red features, respectively, supporting their correspondence with the experimentally observed electronic states. This agreement strongly supports surface stacking as the structural degree of freedom underlying the switching.

Having associated the two electronic states with distinct surface stackings, we next examine the reversibility and nonvolatility of the photo-induced metallic state (Fig. 3). Tracking the electronic structure during a temperature cycle reveals that heating the metallic state (Fig. 3b) to ~ 150 K, followed by cooling to 68 K, fully restores the surface-insulating state (Fig. 3c). A direct comparison of the ARPES spectra (Figs. 3a and 3c) and the corresponding density-of-state (DOS) curves (Fig. 3d) demonstrates that the recovered insulating state is nearly indistinguishable from the pristine ground state, underscoring the reversibility of the switching. Temperature-dependent integrated EDCs over the momentum range $k \in (-0.49, 0.16)$/Å (Fig. 3e) reveal a progressive loss of spectral weight near the Fermi level upon heating, consistent with the reopening of the surface-insulating gap. A quantitative analysis in Fig. 3f yields a characteristic recovery temperature of ~ 122 K via a sigmoid fit, indicating substantial thermal stability of the photo-induced metallic state. Beyond thermal recovery, the photo-induced metallic state can also be optically erased using a lower-fluence pulse sequence (Fig. 3h), establishing fully optical control of the surface electronic structure (see Fig. S3 in Supplementary Information). Repeated write-erase cycles further demonstrate that the switching is highly reproducible (Fig. 3i). Together, these results establish nonvolatile and bidirectional optical control of the surface electronic state.

To clarify how surface stacking affects the spatial distribution and coupling of the low-energy electronic states, we compared the layer-resolved electronic structures and real-space charge distributions for the A- and L-surface configurations (Fig. 4). Figures 4a and 4b show the calculated band structures and layer-resolved DOS within the $\sqrt{13}\times\sqrt{13}$ CDW supercell Brillouin zone. In both cases, the states near the Fermi level are dominated by Ta 5*d* orbitals (see Figs. S5 and S6 in Supplementary Information). For the A-surface configuration, the two outermost layers are gapped, while deeper layers retain a finite density of states at the Fermi level, indicating an insulating surface above a metallic interior. By contrast, the L-surface exhibits finite density of states at the Fermi level throughout the slab, with metallic states extending from the interior to the outermost layer. The A-surface therefore hosts an electronic boundary separating the gapped outer layers from the metallic interior. In the L-surface, this internal boundary is removed as the metallic state extends continuously to the outermost layer. Additionally, we further calculated the partial charge densities between the two outermost layers for both configurations. In the A-surface configuration, the vertically aligned SOD clusters confine the electronic density largely within individual clusters (Fig. 4c). In the L-surface configuration, the lateral stacking offset redistributes the density toward regions connecting neighbouring clusters (Fig. 4d). This geometry introduces interlayer-mediated hopping with a finite in-plane component, enhancing inter-SOD coupling and promoting in-plane electronic delocalization (Figs. 4e and 4f). Such coupling is consistent with the increased in-plane bandwidth of the Ta 5*d* band in the photo-induced metallic state (Figs. 2b and 2d). Together, these results link surface stacking to both the depth distribution and lateral coupling of the low-energy states, providing a microscopic interpretation of the optically driven reconfiguration of the surface-bulk electronic boundary.

## Discussion

These results suggest a microscopic scenario for the laser-induced switching from an insulating to a metallic surface state in 1*T*-$TaSe_2$. Transient suppression of the CDW order by a femtosecond pulse[37] reduces the distinctions among the stacking registries defined by the CDW reconstruction. This non-equilibrium condition may reshape the effective stacking energy landscape, opening a relaxation pathway from the A-surface configuration toward the energetically competitive L-surface configuration. Because CDW stacking reflects the relative alignment of the CDW distortion patterns, this reconfiguration may occur during CDW recovery without requiring large-scale rigid sliding of entire atomic layers. The persistence of the photo-induced metallic state below approximately 120 K is consistent with the metastability of the L-surface configuration.

The microscopic origin of the insulating surface state in 1*T*-$TaSe_2$ nevertheless remains debated, particularly the relative and potentially intertwined roles of dimerization between the two outermost layers and electronic correlations in gap formation[29–31,33–35]. This uncertainty does not alter our central experimental result, namely optical insulator-to-metal switching at the surface of a material with a metallic bulk. If electronic correlations contribute substantially to the surface gap (*e.g.*, a Mott insulating state), our result also suggests a route to persistent and reversible optical control of correlated surface electronic states.

In conclusion, we demonstrate nonvolatile and reversible optical insulator-to-metal switching at the surface of 1*T*-$TaSe_2$, whose bulk is metallic. Comparison with first-principles calculations associates the two electronic states with distinct surface CDW stackings and supports a microscopic picture in which interlayer registry effectively shapes the spatial distribution and coupling of low-energy states. Our results establish persistent optical control of surface metallicity and open a route to reconfiguring the effective surface-bulk electronic boundary in layered quantum materials.

## Materials and Methods

### Sample synthesis and growth.

Stoichiometric amounts of Ta and Se powders were thoroughly mixed and ground in an agate mortar. The homogenized mixture was loaded into a quartz ampoule together with a small amount of iodine as the transport agent. The ampoule was then evacuated, sealed, and placed in a two-zone furnace for chemical vapor transport. Crystal growth was carried out with the hot end maintained at 1100 °C and the cold end at 1050 °C for one week. After cooling to room temperature, pale-yellow, phase-pure single crystals were obtained.

### Laser excitation control and photoemission setup.

The control laser pulses were generated by a Yb-doped fiber laser system (photon energy 1.2 eV, pulse duration ~ 80 fs, spot size ~ 100 μm). The repetition rate was tunable between 1 Hz and 1 MHz. An optical shutter synchronized to the laser output enabled the selection of either single pulses or pulse trains, which were used for writing and erasing operations. ARPES measurements were performed under ultra-high vacuum conditions (better than $3\times10^{-11}$ mbar) using a hemispherical analyser (Scienta Omicron DA30-L) combined with a helium discharge lamp and a vacuum-ultraviolet (VUV) laser source[38]. The He I$\alpha$ line (21.2 eV) delivered an energy resolution better than 5 meV with a beam spot of ~ 1 mm. To ensure homogeneous excitation across this large probe area, the entire sample surface (~ 2 mm×1.5 mm) was scanned by the control laser in both horizontal and vertical directions with a 20 μm step size. By contrast, the VUV laser (7.2 eV, ~ 10 meV resolution, ~ 20 μm spot size) provided localized probing. In this configuration, the overlap of control and probe beams was carefully optimized to guarantee that the detected spectra reflected the photo-induced phase transition. Sample manipulation was carried out using a 6-axis manipulator equipped with temperature control from 6 to 340 K. Bulk single-crystal samples were cleaved in situ at low temperature by knocking off a top-mounted post fixed with silver epoxy, thereby exposing fresh and clean surfaces suitable for ARPES measurements under ultra-high vacuum.

### Density functional theory (DFT) calculations.

The DFT calculations were performed using the *Quantum ESPRESSO* (QE) software package with the generalized gradient approximation (GGA)[39,40]. The Tkatchenko-Scheffler scheme was used to describe van der Waals (vdW) interaction in the $TaSe_2$ system[41]. The plane-wave basis set cut-off energy was set to 320 eV, and a *k*-point grid of 4×4×1 was used to simulate the $\sqrt{13}\times\sqrt{13}$ unit cell of the star-of-David. The DFT+U scheme was also used to describe the electronic properties of the $TaSe_2$ system[36], and the Coulomb correlation of Hubbard was set to U = 2 eV. A vacuum layer of approximately 15 Å was introduced in the vertical direction to avoid artificial interactions from periodic boundary conditions. All atomic structures were fully relaxed. The convergence criterion for energy was $10^{-6}$ eV, and the forces on all atoms were less than 0.02 eV/Å. The optimized lattice constants for the A-surface and L-surface structures were found to be a = b = 12.50 Å, which closely matched previous experimental observations of a = b = 12.54 Å[42].

**Acknowledgements:** This work was supported by the National Key R&D Program of China (grant nos. 2024YFA1408400, 2022YFA1405600), the National Natural Science Foundation of China (grant nos. 12204042, 12674002, 12274024, 12321004, 52522201, 52250308), the Beijing Natural Science Foundation (grant no. Z240005), and the Beijing Institute of Technology Research Fund Program for Young Scholars. J.L. acknowledges funding from the Alexander von Humboldt Foundation.

**Author contributions:** X.S. conceived the project. J.L. and P.L. performed the ARPES experiments with assistance from M.P., Y.Z., F.C., Y.C., Z.X., T.Q. and X.S. H.L. and T.Y. grew the single crystals. L.S., L.H. and Y.Y. performed the theoretical calculations and analysis. J.L. analyzed the data and prepared the figures with help from L.S. and X.S. J.L., S.M., and X.S. wrote the manuscript with contributions from all authors.

**Data availability:** The data that support the findings of this study are available from the corresponding author upon reasonable request.

**Competing interests:** The authors declare no competing interests.

**References:**

1. Keimer, B. & Moore, J. E. The physics of quantum materials. *Nat. Phys.* **13**, 1045–1055 (2017).

2. Geim, A. K. & Grigorieva, I. V. Van der Waals heterostructures. *Nature* **499**, 419–425 (2013).

3. Novoselov, K. S., Mishchenko, A., Carvalho, A. & Neto, A. H. C. 2D materials and van der Waals heterostructures. *Science* **353**, aac9439 (2016).

4. Hasan, M. Z. & Kane, C. L. Colloquium: Topological insulators. *Rev. Mod. Phys.* **82**, 3045–3067 (2010).

5. Lee, J. J. *et al.* Interfacial mode coupling as the origin of the enhancement of Tc in FeSe films on $SrTiO_3$. *Nature* **515**, 245–248 (2014).

6. Liu, C. *et al.* Two-dimensional superconductivity and anisotropic transport at $KTaO_3$(111) interfaces. *Science* **371**, 716–721 (2021).

7. Sipos, B. *et al.* From Mott state to superconductivity in 1T-$TaS_2$. *Nat. Mater.* **7**, 960–965 (2008).

8. Chen, Y. *et al.* Strong correlations and orbital texture in single-layer 1T-$TaSe_2$. *Nat. Phys.* **16**, 218–224 (2020).

9. Hansmann, P., Ayral, T., Vaugier, L., Werner, P. & Biermann, S. Long-Range Coulomb Interactions in Surface Systems: A First-Principles Description within Self-Consistently Combined $GW$ and Dynamical Mean-Field Theory. *Phys. Rev. Lett.* **110**, 166401 (2013).

10. Yasuda, K., Wang, X., Watanabe, K., Taniguchi, T. & Jarillo-Herrero, P. Stacking-engineered ferroelectricity in bilayer boron nitride. *Science* **372**, 1458–1462 (2021).

11. Cao, Y. *et al.* Unconventional superconductivity in magic-angle graphene superlattices. *Nature* **556**, 43–50 (2018).

12. Cao, Y. *et al.* Correlated insulator behaviour at half-filling in magic-angle graphene superlattices. *Nature* **556**, 80–84 (2018).

13. Fausti, D. *et al.* Light-Induced Superconductivity in a Stripe-Ordered Cuprate. *Science* **331**, 189–191 (2011).

14. Rohwer, T. *et al.* Collapse of long-range charge order tracked by time-resolved photoemission at high momenta. *Nature* **471**, 490–493 (2011).

15. Radu, I. *et al.* Transient ferromagnetic-like state mediating ultrafast reversal of antiferromagnetically coupled spins. *Nature* **472**, 205–208 (2011).

16. Sie, E. J. *et al.* An ultrafast symmetry switch in a Weyl semimetal. *Nature* **565**, 61–66 (2019).

17. Stojchevska, L. *et al.* Ultrafast Switching to a Stable Hidden Quantum State in an Electronic Crystal. *Science* **344**, 177–180 (2014).

18. Zeng, Z. *et al.* Photo-induced nonvolatile rewritable ferroaxial switching. *Science* **390**, 195–198 (2025).

19. Liu, Q. M. *et al.* Photoinduced multistage phase transitions in $Ta_2NiSe_5$. *Nat. Commun.* **12**, 2050 (2021).

20. Liu, Q. *et al.* Room-temperature non-volatile optical manipulation of polar order in a charge density wave. *Nat. Commun.* **15**, 8937 (2024).

21. Gao, F. Y. *et al.* Snapshots of a light-induced metastable hidden phase driven by the collapse of charge order. *Sci. Adv.* **8**, eabp9076 (2022).

22. Sun, K. *et al.* Hidden CDW states and insulator-to-metal transition after a pulsed femtosecond laser excitation in layered chalcogenide 1T-$TaS_{2-x}Se_x$. *Sci. Adv.* **4**, eaas9660 (2018).

23. Stahl, Q. *et al.* Collapse of layer dimerization in the photo-induced hidden state of 1T-$TaS_2$. *Nat. Commun.* **11**, 1247 (2020).

24. Maklar, J. *et al.* Coherent light control of a metastable hidden state. *Sci. Adv.* **9**, eadi4661 (2023).

25. Liu, J. *et al.* Nonvolatile optical control of interlayer stacking order in 1T-$TaS_2$. *npj Quantum Mater.* **11**, 6 (2026).

26. Gierster, L., Vempati, S. & Stähler, J. Ultrafast generation and decay of a surface metal. *Nat. Commun.* **12**, 978 (2021).

27. Duan, S. *et al.* Optical manipulation of electronic dimensionality in a quantum material. *Nature* **595**, 239–244 (2021).

28. Aiura, Y. *et al.* Electronic structure of layered 1T-$TaSe_2$ in commensurate charge-density-wave phase studied by angle-resolved photoemission spectroscopy. *Phys. Rev. B* **68**, 073408 (2003).

29. Perfetti, L. *et al.* Spectroscopic Signatures of a Bandwidth-Controlled Mott Transition at the Surface of 1T-$TaSe_2$. *Phys. Rev. Lett.* **90**, 166401 (2002).

30. Colonna, S. *et al.* Mott Phase at the Surface of 1T-$TaSe_2$ Observed by Scanning Tunneling Microscopy. *Phys. Rev. Lett.* **94**, 036405 (2004).

31. Lee, S.-H. & Cho, D. Surface-Reconstruction-Driven Insulating Behavior in Metallic Charge-Density-Wave 1T-$TaSe_2$. *Nano Lett.* (2026).

32. Ren, Y. J. *et al.* Large variation of interlayer coupling and electron hopping in 1T-$TaSe_2$ resolved by angle-resolved photoemission spectroscopy. *Phys. Rev. B* **112**, 115123 (2025).

33. Straub, M. *et al.* Nature of Metallic and Insulating Domains in the Charge-Density-Wave System 1T-$TaSe_2$. *Phys. Rev. Lett.* **135**, 096501 (2025).

34. Mignani, N. *et al.* Evidence of Surface Interlayer Dimerization in the Commensurate Charge Density Wave Phase of 1T-$TaSe_2$. *Phys. Rev. Lett.* **135**, 256203 (2025).

35. Tian, N. *et al.* Dimensionality-driven metal to Mott insulator transition in two-dimensional 1T-$TaSe_2$. *Natl. Sci. Rev.* **11**, nwad144 (2024).

36. Wang, W., Zhao, B., Ming, X. & Si, C. Multiple Quantum States Induced in 1T-$TaSe_2$ by Controlling the Stacking Order of Charge Density Waves. *Adv. Funct. Mater.* **33**, 2214583 (2023).

37. Shi, X. *et al.* Ultrafast electron calorimetry uncovers a new long-lived metastable state in 1T-$TaSe_2$ mediated by mode-selective electron-phonon coupling. *Sci. Adv.* **5**, eaav4449 (2019).

38. Pan, M. *et al.* Time-resolved ARPES with probe energy of 6.0/7.2 eV and switchable resolution configuration. *Rev. Sci. Instrum.* **95**, 013001 (2024).

39. Giannozzi, P. *et al.* QUANTUM ESPRESSO: a modular and open-source software project for quantum simulations of materials. *J. Phys.: Condens. Matter* **21**, 395502 (2009).

40. Giannozzi, P. *et al.* Advanced capabilities for materials modelling with Quantum ESPRESSO. *J. Phys.: Condens. Matter* **29**, 465901 (2017).

41. Tkatchenko, A. & Scheffler, M. Accurate Molecular Van Der Waals Interactions from Ground-State Electron Density and Free-Atom Reference Data. *Phys. Rev. Lett.* **102**, 073005 (2008).

42. Brouwer, R. & Jellinek, F. The low-temperature superstructures of 1T-$TaSe_2$ and 2H-$TaSe_2$. *Phys. B+C* **99**, 51–55 (1980).

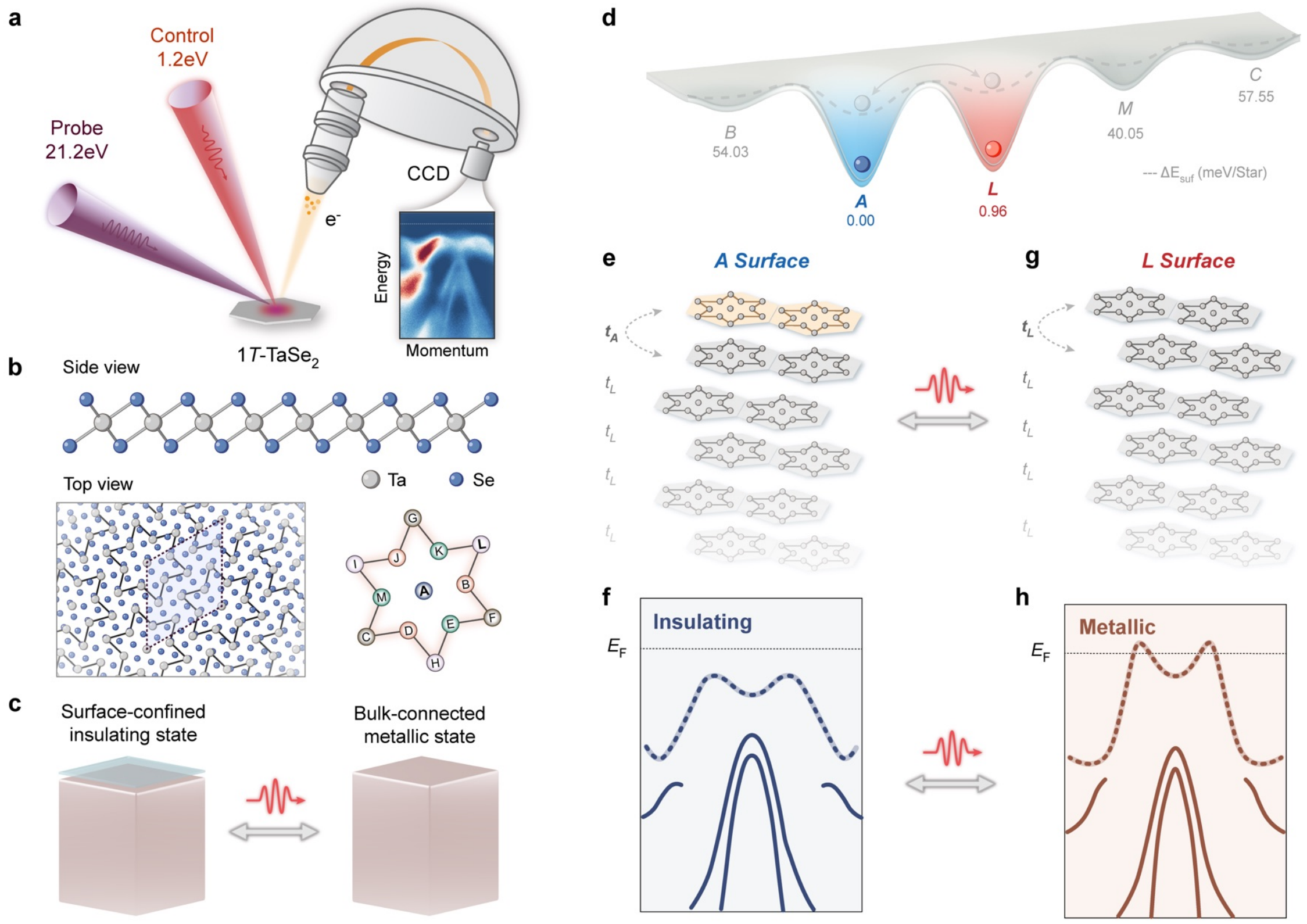

**Fig. 1 | Schematic illustration of the laser-induced switching. a,** Experimental setup of the laser excitation and ARPES measurement. **b,** Crystal structure of 1T-$TaSe_2$. Top: side view. Bottom left: top view, where 13 Ta atoms form a star-of-David (SOD) cluster due to CDW reconstruction. The dashed box marks the $\sqrt{13}\times\sqrt{13}$ superlattice. Bottom right: site labelling scheme for the 13 Ta atoms within a cluster. **c,** Schematic illustration of reversible optical switching between a surface-confined insulating state and a bulk-connected metallic state. **d,** Schematic energy landscape of different surface-stacking configurations. The numbers denote the calculated relative surface energies in meV/Star, referenced to the A-surface configuration. **e, g,** Surface-stacking configurations assigned to the equilibrium insulating state and the photo-induced metallic state, respectively. In the A-surface configuration (**e**), the top two layers adopt the A stacking (site A to site A alignment of SOD in adjacent layers), while deeper layers follow the L stacking (site A to site L alignment of SOD in adjacent layers). In the L-surface configuration (**g**), the top two layers adopt the L stacking, consistent with the bulk. **f, h,** Corresponding schematic low-energy band structures of the A- and L-surface configurations, illustrating the surface-gapped insulating state (**f**) and the metallic state with bands crossing the Fermi level and a broadened Ta 5d dispersion (**h**).

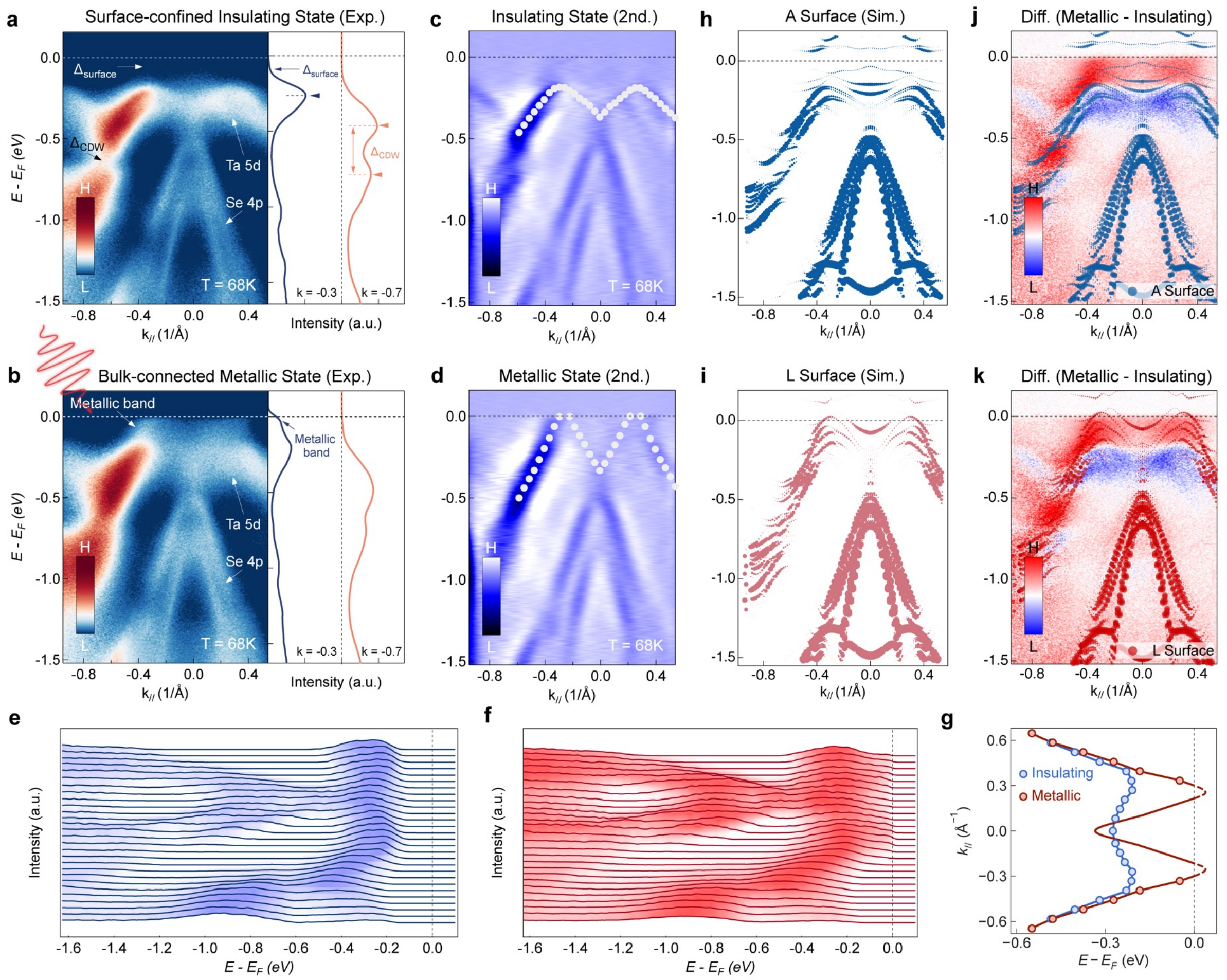

**Fig. 2 | Nonvolatile photo-induced reconstruction of the electronic structure. a,** ARPES spectra of the surface-confined insulating state along the Γ-M direction measured with a He-Iα source (21.2 eV) at 68 K. The right panels show EDCs at k = -0.3/Å and k = -0.7/Å, highlighting the surface-insulating gap $\Delta_{\mathrm{Surface}}$ and the CDW gap $\Delta_{\mathrm{CDW}}$, respectively. **b,** Similar to **a**, but for the photo-induced bulk-connected metallic state, revealing the emergence of a metallic band and filling of the CDW gap. **c, d,** Second-derivative spectra corresponding to **a** and **b**, respectively. White circles mark the low-energy band dispersion near the Fermi level. **e, f,** Intensity maps of the spectra for the insulating (blue) and metallic state (red) with corresponding EDC series. **g,** Comparison of the extracted low-energy dispersion for the insulating (blue) and metallic (red) states. Solid lines are guides to the eye, highlighting the enhanced in-plane dispersion in the photo-induced metallic state. **h, i,** DFT+U band structures calculated for the A-surface (**h**) and the L-surface (**i**), corresponding to the insulating and metallic state, respectively. **j, k,** Differential maps between the experimental spectra of the metallic and insulating states, highlighting the redistribution of spectral weight and the accompanying reconstruction of the low-energy bands. Blue and red represent spectral-weight loss and gain, respectively. The calculated dispersions of the A-surface (**h**, blue) and L-surface (**i**, red) are overlaid on **j** and **k**, respectively.

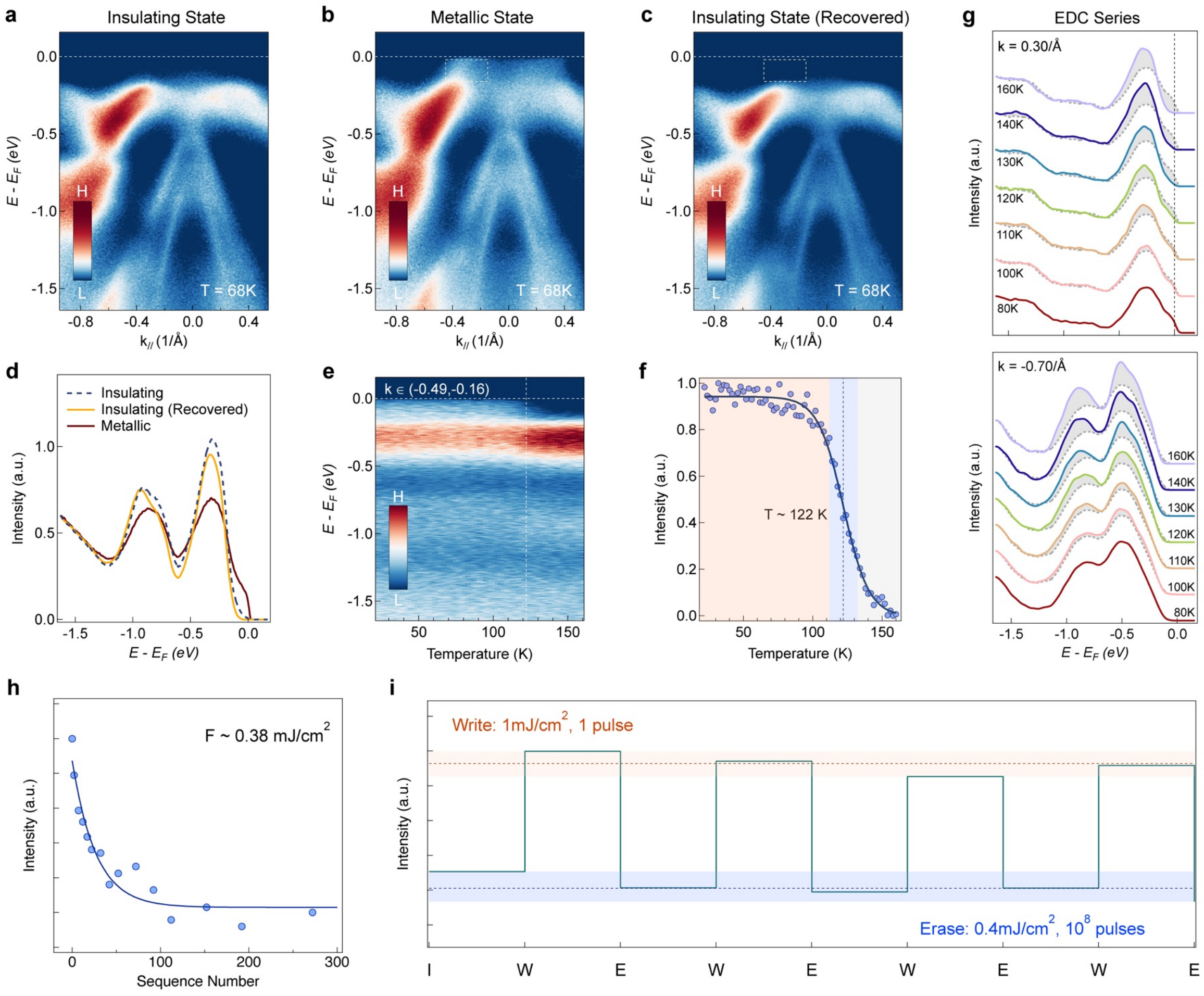

**Fig. 3 | Thermal recovery and bidirectional optical programming of the metallic state. a-c,** ARPES spectra along the direction Γ-M measured with a He-Iα source (21.2 eV) at 68 K, showing the initial insulating ground state (**a**), the photo-induced metallic state (**b**), and the recovered insulating state after a thermal cycle (**c**). **d,** Corresponding DOS curves extracted from **a-c**. **e,** Temperature-dependent integrated EDCs over the momentum range k ∈ (-0.49, 0.16)/Å, showing the progressive suppression of spectral intensity near $E_F$ as the system returns to the insulating state upon heating. The vertical dashed line marks the characteristic recovery temperature. **f,** Temperature evolution of the spectral weight within the white dashed box in **b** and **c**. The data are fitted with a sigmoid function (blue line), yielding a transition temperature of T ~ 122 K. The shaded blue region denotes the half-width at half-maximum extracted from the fit. **g,** EDC series at k = 0.3/Å (top) and k = -0.7/Å (bottom) at different temperatures, showing the gradual recovery of the insulating state upon heating. **h,** Evolution of the integrated near-$E_F$ spectral weight during optical erasure using laser pulse sequences, where each sequence contains 500,000 pulses at a fixed fluence of 0.38 mJ/cm$^2$. **i,** Repeated optical write-erase cycles. The transition to the metallic state is triggered by a single laser pulse (1 mJ/cm$^2$, 1 pulse), while the insulating state is restored using a pulse sequence (0.4 mJ/cm$^2$, $10^8$ pulses).

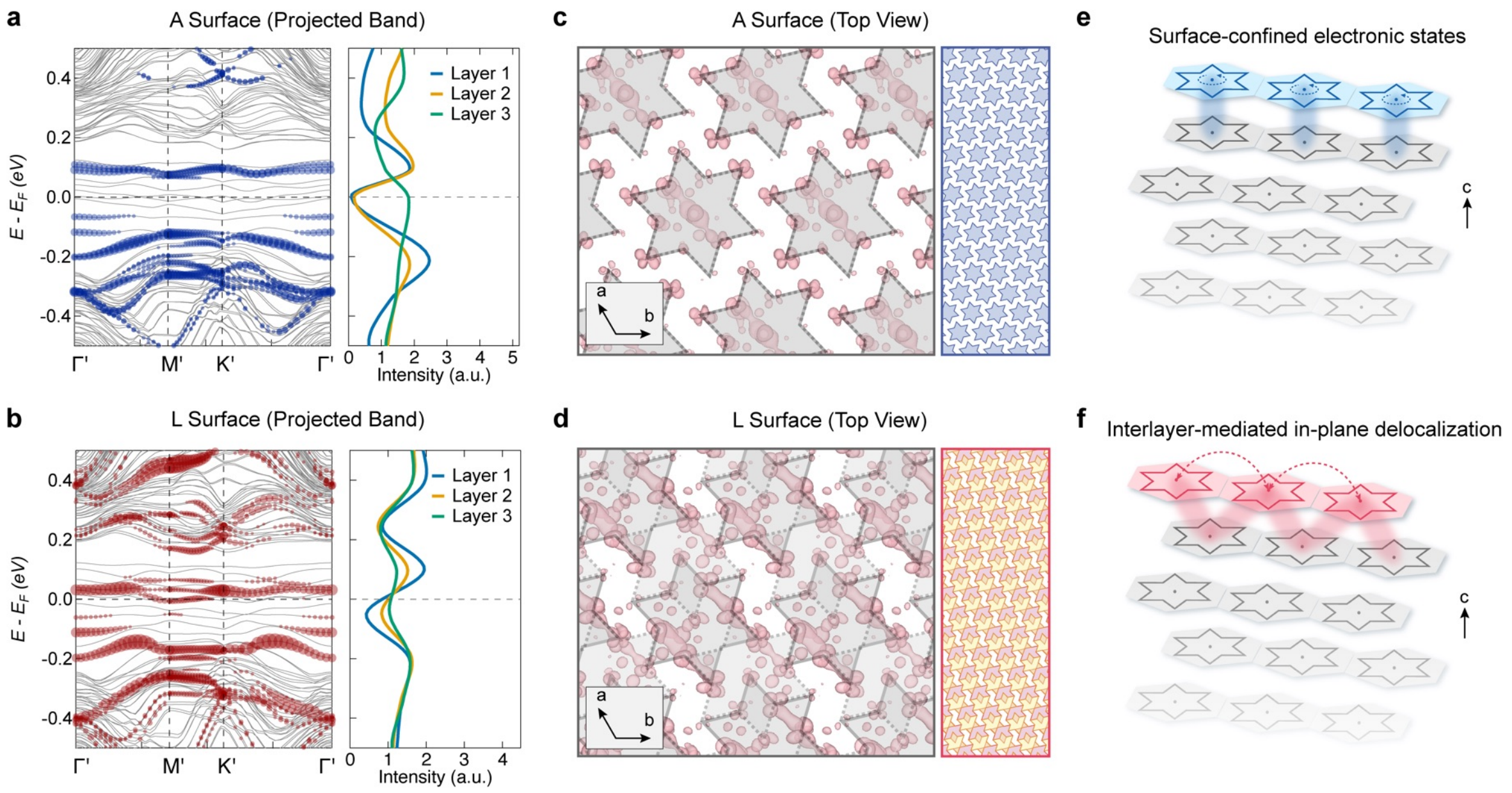

**Fig. 4 | Stacking-controlled electronic boundary and surface-bulk connectivity. a, b,** Calculated layer-resolved electronic structures of 6-layer slabs terminated by the A-surface (**a**) and the L-surface (**b**). The momentum-space high-symmetry points are defined with respect to the Brillouin zone of the CDW superlattice. In **a**, the blue spectral intensity represents the contribution of the central Ta $5d_{z^2}$ orbital in the SOD cluster of the topmost layer. In **b**, the corresponding spectral weight is shown in red. The panels on the right show the projected density of states of the same central Ta $5d_{z^2}$ orbital in the first (blue), second (orange) and third (green) layers, respectively. **c, d,** In-plane partial charge-density distributions for the A-surface (**c**) and the L-surface (**d**). The SOD clusters illustrate the reconstructed CDW pattern and indicate the stacking arrangement of the top two layers. The right-hand panels provide enlarged top-view schematics that illustrate the corresponding stacking arrangements over a wider range. **e, f,** Schematic illustration of the stacking-dependent electronic connectivity.

Supplementary Information for

# Nonvolatile optical switching of surface metallicity in 1*T*-$TaSe_2$

## Section I. Band dispersion and line-shape analysis of ARPES spectra.

To clarify the evolution of the electronic structure during the laser-induced switching from a surface-confined insulating state to a bulk-connected metallic state, we performed a detailed line-shape analysis of the ARPES spectra. As shown in Fig. S1, the DOS-normalized spectra and the corresponding momentum-distribution curves reveal that the most pronounced reconstruction occurs in the Ta 5*d*-derived states near the Fermi level. Following photoexcitation, the low-energy Ta 5*d* states acquire substantially enhanced in-plane dispersion and evolve into a metallic band crossing the Fermi level. The extracted low-energy band dispersions are overlaid on the DOS-normalized spectra in Figs. S1a and S1c, while the corresponding MDC series are shown in Figs. S1b and S1d to illustrate the evolution of the spectral line shape. Together, these analyses demonstrate that optical switching reconstructs the low-energy Ta 5*d* electronic structure, leading to enhanced electronic itinerancy and the formation of a more dispersive metallic band.

## Section II. All-optical bidirectional switching probed by VUV laser source.

To further verify the robustness of the observed optical switching, we performed complementary ARPES measurements using a VUV probe with a photon energy of 7.2 eV. Although the photoemission matrix elements at this photon energy strongly suppress the spectral weight of the Ta 5*d*-derived states relative to the Se p bands (Figs. S2a, S2b, S2d and S2e), the corresponding EDCs still resolve the surface gap in the insulating state and the emergence of metallic spectral weight following photoexcitation (Figs. S2c and S2f). These features are consistent with those measured using the He-lamp source, confirming that the spectral reconstruction is independent of the probe configuration. The small spot size of the VUV probe further allows spatial overlap with the 1.2-eV control beam, enabling localized optical excitation and detection.

As shown in Fig. S2g, increasing the excitation fluence produces a systematic redistribution of spectral weight and a reconstruction of the low-energy bands from the surface-confined insulating state towards the photo-induced metallic state. Although the Ta 5*d* bands are less prominent at this photon energy, the spectral contrast remains sufficient to quantify the switching. By tracking the evolution of EDC peak positions (Fig. S2h) and spectral intensities (Fig. S2i), we identify a consistent fluence-dependent trend, yielding a critical threshold of $\sim 0.7\ \mathrm{mJ/cm^2}$. Furthermore, as shown in Fig. S2j, the switching is mainly governed by excitation fluence rather than the number of incident pulses, excluding the possibility of cumulative heating effects.

The photo-induced metallic state can also be optically erased. As shown in Figs. S3a–c, excitation with an appropriate pulse sequence restores an ARPES spectrum closely resembling that of the initial surface-confined insulating state. The fluence dependence measured at a fixed total pulse number reveals a distinct erasure window, whereas the recovery becomes less effective as the fluence approaches the threshold for writing the metallic state (Fig. S3d). Together with the repeated write-erase cycles shown in Figs. 3h and 3i, these results establish fully optical, bidirectional and reproducible programming of the surface electronic structure.

**Section III. Detailed analysis of calculated electronic structures.**

To clarify the distinct optical responses of 1*T*-$TaSe_2$ and 1*T*-$TaS_2$, we compare their low-energy stacking configurations and corresponding electronic structures in Fig. S4. Despite their closely related CDW structures, the two materials exhibit qualitatively different energy landscapes. In 1*T*-$TaS_2$, the lowest-energy configuration is a bulk-dimerized AL stacking associated with the insulating state, whereas the metallic bulk L stacking constitutes the next-lowest-energy configuration. Optical switching between these competing structures therefore requires a reorganization of the stacking sequence throughout the crystal, giving rise to a bulk insulator-to-metal transition. By contrast, the lowest-energy configuration of 1*T*-$TaSe_2$ is the reconstructed A-surface, in which the A-type stacking is confined to the two outermost layers while the interior retains the metallic L-type stacking. The unreconstructed L-surface, which continues the bulk L-type sequence to the outermost layer, lies only slightly higher in energy. The relevant competition in 1*T*-$TaSe_2$ is therefore localized near the surface, such that optical switching primarily removes the insulating surface termination and connects the pre-existing metallic interior to the surface. These contrasting energy hierarchies explain why photo-induced switching in 1*T*-$TaS_2$ transforms the bulk electronic phase, whereas in 1*T*-$TaSe_2$ it rewrites the electronic boundary and reconfigures surface-bulk connectivity.

Furthermore, to resolve the orbital character and layer distribution of the low-energy states, we calculated the orbital- and layer-projected band structures for the A-surface and L-surface configurations. In both cases, the electronic states within the energy range of [-0.5, 0.5] eV near the Fermi level are dominated by Ta $d_{z^2}$ orbitals, with substantially smaller contributions from the $d_{xy}$, $d_{yz}$, and $d_{xz}$ orbitals (Figs. S5 and S6). Their layer distributions, however, differ markedly. For the A-surface, the two outermost layers exhibit a gap near the Fermi level, whereas the deeper layers retain metallic spectral weight, consistent with an insulating electronic structure confined near the surface above a metallic interior. By contrast, the low-energy metallic states of the L-surface exhibit appreciable weight across all layers, extending continuously from the interior to the outermost layer. This contrast further supports the interpretation of the two configurations as a surface-confined insulating state and a bulk-connected metallic state, respectively.

We further calculated the differential charge transfer of the selected low-energy states for the A-surface and L-surface configurations. In the A-surface structure, the vertically aligned SOD clusters produce charge density that remains predominantly confined within individual clusters (Figs. S7c and S7e). In the L-surface structure, the lateral stacking offset redistributes the low-energy charge density towards regions connecting neighbouring SOD clusters (Figs. S7d and S7f). This redistribution is consistent with a broader network of interlayer-mediated hopping pathways with finite in-plane components, which enhances inter-SOD connectivity and provides a microscopic explanation for the broadened in-plane Ta 5*d* dispersion observed experimentally.

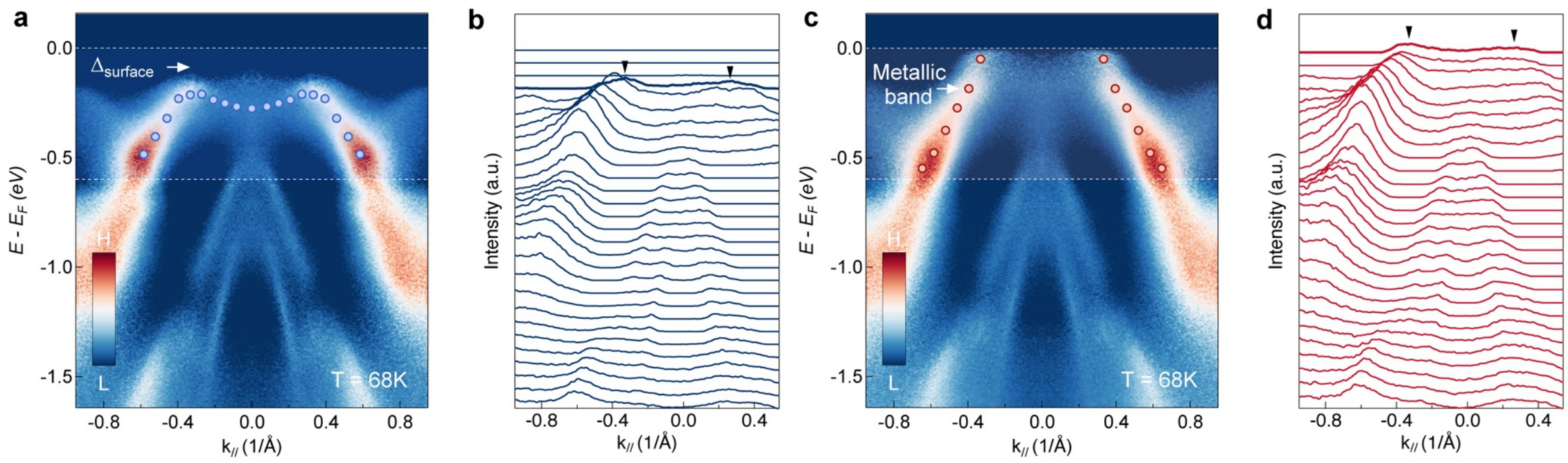

**Fig. S1 | Band dispersion analysis of ARPES spectra. a, c,** DOS-normalized ARPES spectra for the surface-insulating state (**a**) and the photo-induced metallic state (**c**) at 68 K. DOS normalization enhances the visibility of the band positions by suppressing matrix-element effects. Extracted band dispersions obtained by fitting the spectra in (**a**) (blue markers) and (**c**) (red markers) are overlaid. **b, d,** MDC series corresponding to the spectra of the surface-insulating state and the metallic state, respectively. The arrows indicate the peak position of the lowest-energy MDC.

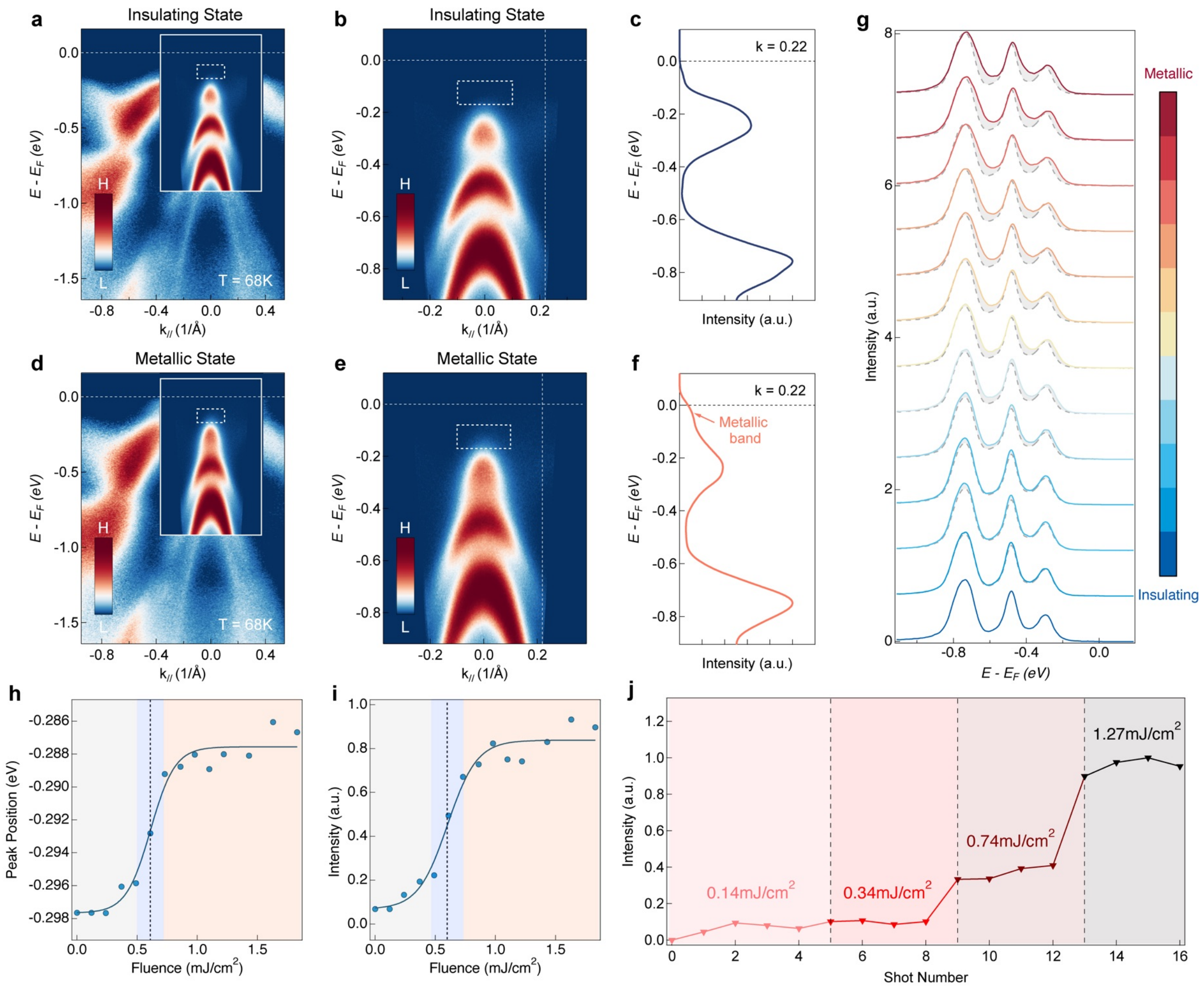

**Fig. S2 | Band evolution measured by VUV laser source. a, d,** ARPES spectra of the surface-insulating state (**a**) and the photo-induced metallic state (**d**) measured with a He-Iα source (21.2 eV). The insets within the white boxes show the corresponding measurements performed with a VUV laser source (7.2 eV). **b, e,** Close-up views of the laser-based spectra corresponding to the regions marked in **a** and **d**, respectively. Due to photoemission cross-section effects, the spectral intensity of the Ta 5*d* band is strongly suppressed in the laser measurements compared to the He lamp results. **c, f,** Corresponding EDCs taken at k = 0.22/Å (white dashed line) from the spectra in the insulating state (**b**) and the metallic state (**e**). **g,** Evolution of normalized EDCs integrated over the momentum range marked by the white dashed box in **b** and **e**. The dashed line indicates the unexcited reference spectrum. **h, i,** Fluence dependence of the peak position near $E-E_F \sim -0.29$ eV (**h**) and of the integrated spectral intensity (**i**) within the white dashed box in (**b**) and (**e**). The solid curves represent sigmoid fits, the vertical dashed line marks the critical fluence threshold, and the shaded blue region denotes the full width at half maximum of the transition. Both analyses yield a nearly identical threshold fluence of 0.7 mJ/cm$^2$. **j,** Evolution of the spectral intensity within the boxed region in (**b**) and (**e**) as a function of laser pulse number and fluence. Different shaded regions correspond to measurements taken at different fluence levels, as labelled. The results demonstrate that the transition is governed primarily by the excitation fluence, while the accumulated pulse number has little influence.

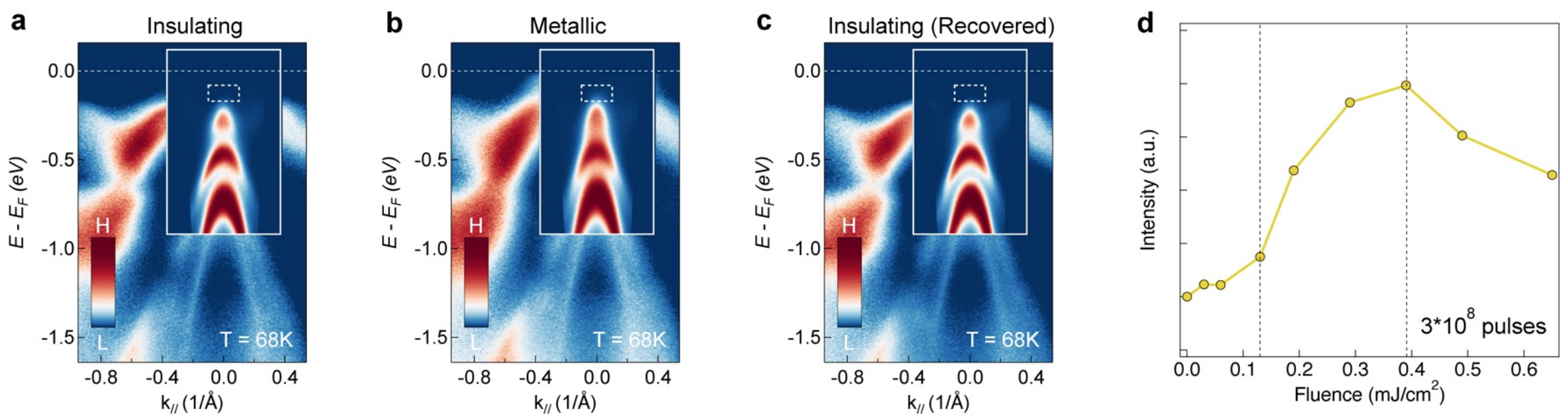

**Fig. S3 | Optical erasure of the photo-induced metallic state. a-c,** ARPES spectra of the initial surface-insulating state (**a**), the photo-induced metallic state (**b**), and the recovered insulating state after excitation by a laser pulse sequence (**c**) measured at 68 K. **d,** Fluence dependence of the spectral intensity within the white dashed box in **a-c**, measured with a fixed total number of pulses ($3\times10^8$).

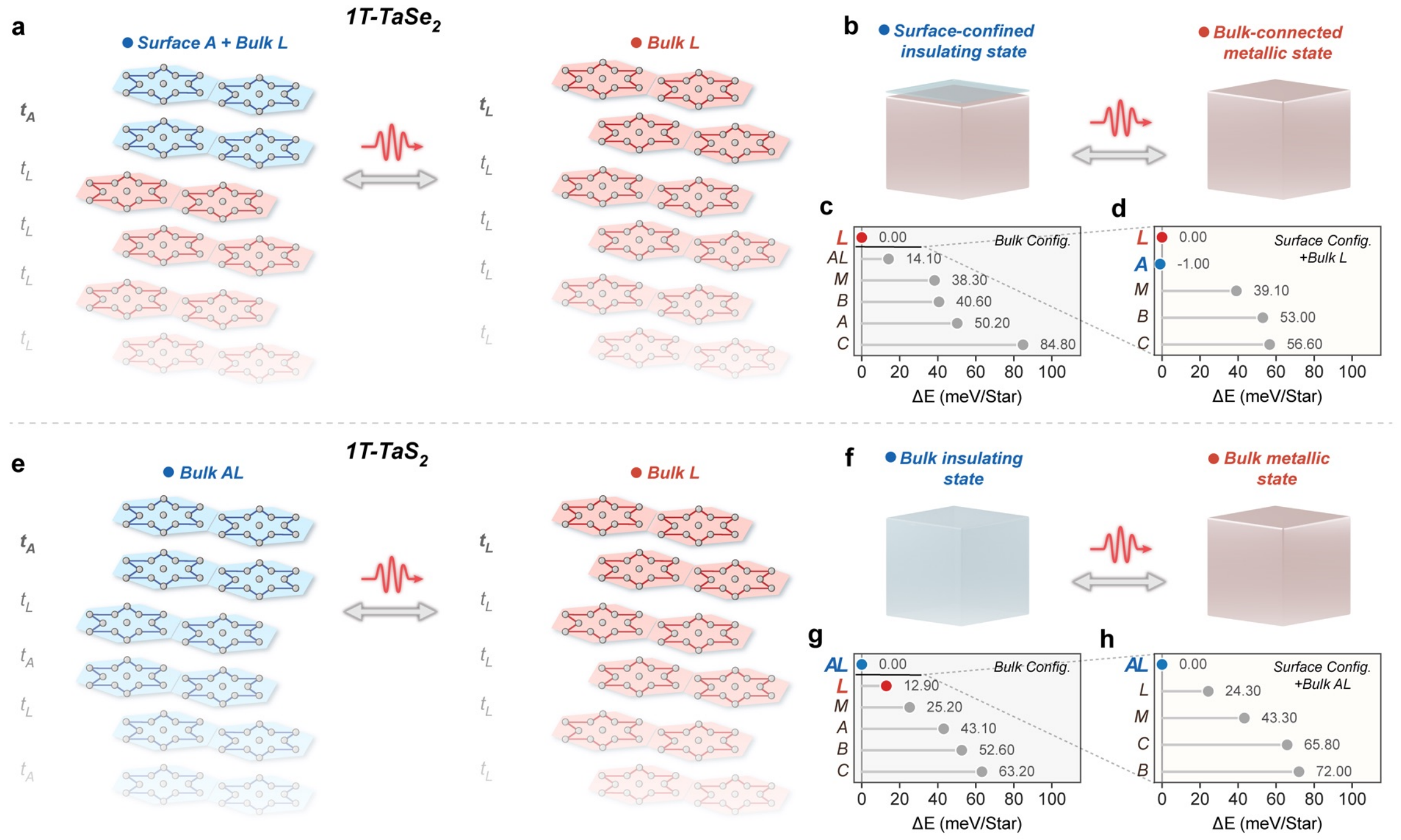

**Fig. S4 | Comparison of stacking-controlled optical switching in 1*T*-$TaSe_2$ and 1*T*-$TaS_2$. a, e,** Schematic interlayer-stacking configurations involved in the light-induced phase transition in 1*T*-$TaSe_2$ (**a**) and 1*T*-$TaS_2$ (**e**), respectively. **b, f,** Schematic illustration of the laser switching in 1*T*-$TaSe_2$ (**b**) and 1*T*-$TaS_2$ (**f**), respectively. Optical switching in 1*T*-$TaSe_2$ converts a surface-confined insulating state into a bulk-connected metallic state, whereas that in 1*T*-$TaS_2$ corresponds to a bulk insulator-to-metal transition. **c, g,** Calculated relative energies of different bulk stacking configurations in 1*T*-$TaSe_2$ (**c**) and 1*T*-$TaS_2$ (**g**). **d, h,** Calculated relative energies of different surface stacking configurations in 1*T*-$TaSe_2$ (**d**) and 1*T*-$TaS_2$ (**h**). The bulk stacking is fixed to its respective lowest-energy configuration (L for 1*T*-$TaSe_2$ and AL for 1*T*-$TaS_2$). In **c**, **d**, **g**, and **h**, the blue and red symbols highlight the two lowest-energy stacking configurations illustrated in **a** and **e**, respectively.

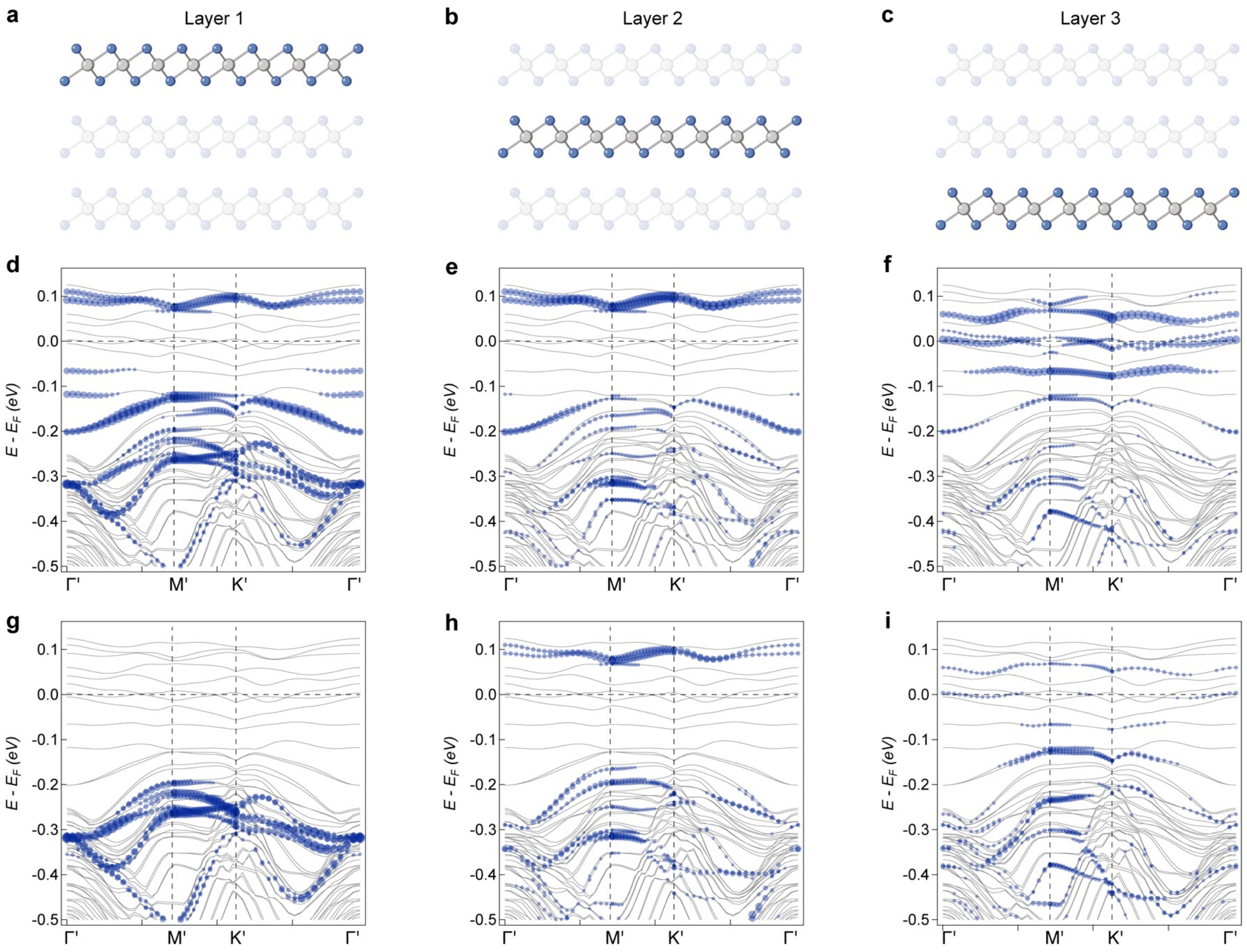

**Fig. S5 | Layer-resolved orbital-projected band structures for the A-surface configuration. a-c,** Structural schematics highlighting different layers: the first layer (**a**), the second layer (**b**), and the third layer (**c**). **d-f,** Corresponding Ta $d_{z^2}$-orbital-projected band structures of 6-layer simulations for the layers shown in **a-c**. **g-i,** Corresponding Ta $d_{xy}$+$d_{yz}$+$d_{xz}$-orbital-projected band structures of 6-layer simulations for the layers shown in **a-c**.

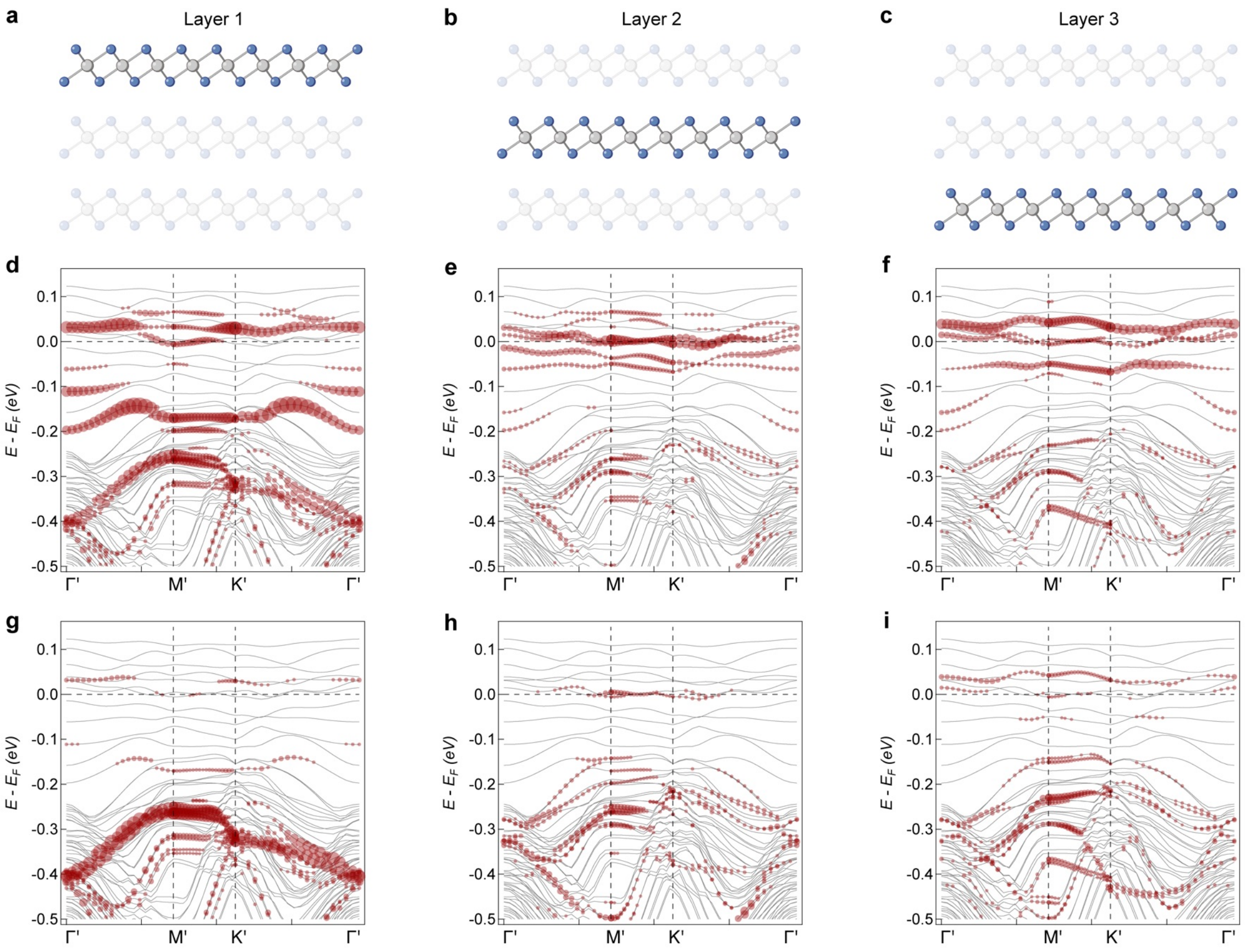

**Fig. S6 | Layer-resolved orbital-projected band structures for the L-surface configuration. a-c,** Structural schematics highlighting different layers: the first layer (**a**), the second layer (**b**), and the third layer (**c**). **d-f,** Corresponding Ta $d_{z^2}$-orbital-projected band structures of 6-layer simulations for the layers shown in **a-c**. **g-i,** Corresponding Ta $d_{xy}+d_{yz}+d_{xz}$-orbital-projected band structures of 6-layer simulations for the layers shown in **a-c**.

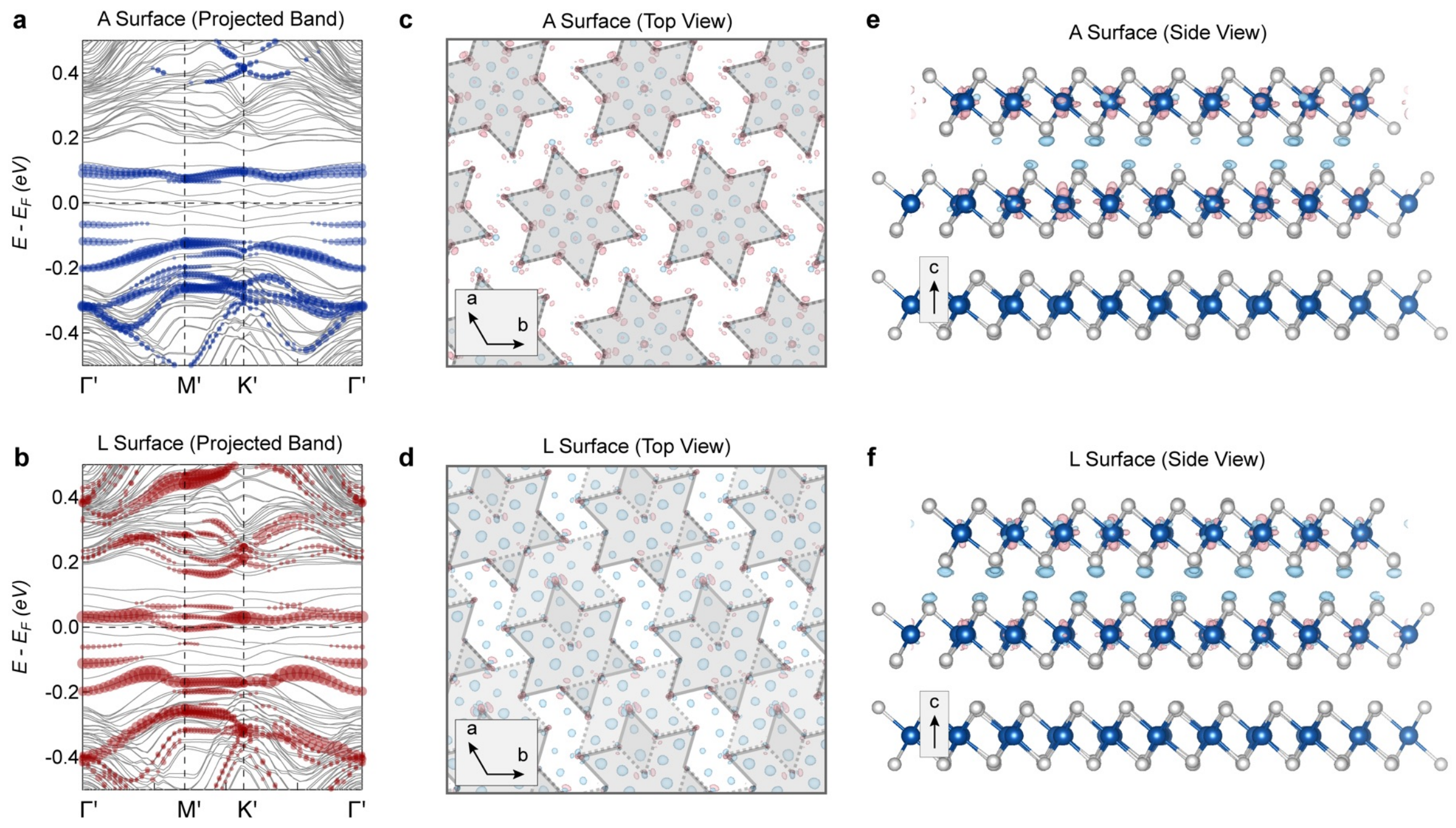

**Fig. S7 | Simulated differential charge transfer distributions for different surface stackings. a, b,** Calculated layer-resolved electronic structures of 6-layer slabs terminated by the A-surface (**a**) and the L-surface (**b**). The momentum-space high-symmetry points are defined with respect to the Brillouin zone of the CDW superlattice. **c, d,** In-plane partial charge transfer distributions for the A-surface (**c**) and the L-surface (**d**). The SOD clusters illustrate the reconstructed CDW pattern and indicate the stacking arrangement of the top two layers. **e, f,** Out-of-plane charge transfer distributions for the A-surface (**e**) and the L-surface (**f**). The direction perpendicular to the page corresponds to the in-plane A-L site connection.